# Hybrid density functional study of band alignment in ZnO/GaN and ZnO/(Ga$_{1-x}$Zn$_x$)(N$_{1-x}$O$_x$)/GaN heterostructures

Zhenhai Wang, Mingwen Zhao [a)], Xiaopeng Wang, Yan Xi, Xiujie He, Xiangdong Liu, and Shishen Yan

*School of Physics and State Key Laboratory of Crystal Materials, Shandong University, Jinan, Shandong, 250100, China*

The band alignment in ZnO/GaN and related heterostructures are crucial for the uses in solar harvesting technology. Here, we report our density functional calculations of the band alignment and optical properties of ZnO/GaN and ZnO/(Ga$_{1-x}$Zn$_x$)(N$_{1-x}$O$_x$)/GaN heterostructures using a Heyd-Scuseria-Ernzerhof (HSE) hybrid functional. We found that the conventional GGA functionals underestimate not only the band gap but also the band offset of these heterostructures. Using the hybrid functional calculations, we show that the (Ga$_{1-x}$Zn$_x$)(N$_{1-x}$O$_x$) solid solution has a direct band gap of about 2.608 eV, in good agreement with the experimental data. More importantly, this solid solution forms type-II band alignment with the host materials. A GaN/(Ga$_{1-x}$Zn$_x$)(N$_{1-x}$O$_x$)/ZnO core-shell solar cell model is presented to improve the visible light adsorption ability and carrier collection efficiency.

Zinc oxide (ZnO) and gallium nitride (GaN) crystals have been attracting much attention in various fields, such as exciton-based optoelectronic devices and solar harvesting, due to their excellent properties.[1, 2] The slight lattice mismatch (~1.86%) between wurtzite (*w*-) ZnO and GaN and the type-II band alignment of the two crystals, has motivated intensive studies on ZnO/GaN heterostructures, because type-II band alignment is always accompanied by natural charge separation which is advantageous for promising solar energy harvesting, such as water splitting, dye-sensitized and even regular solar cells.[3-5] The synthesis of ZnO/GaN core-shell heterostructured nanowires (*h*-NWs) and nanotubes (*h*-NTs) has been reported,[6-8] and the GaN (core)/ZnO (shell) *h*-NTs were found to have absorption spectra in visible light region (1.9-3.6 eV).[7, 8] Based on first-principles calculation and an effective mass model, we predicted that ZnO (core)/GaN (shell) *h*-NWs rather than GaN (core)/ZnO (shell) ones exhibit natural charge spatial separation behaviors in our previous work.[9] Thanks to the unique geometry of core/shell *h*-NWs which have GaN/ZnO interface extending along the axial direction and carrier separation taking place in axial direction of the *h*-NWs, photo-generated carriers can reach the interface with high efficiency without substantial bulk recombination, which improves carrier collection and overall efficiency with respect to comparable axially-modulated *h*-NWs. This provides a novel solar cell structure with high efficiency.

Band offset, a fundamental parameter of ZnO/GaN heterostructures, is crucial for their promising applications in solar harvesting. Considerable efforts have been made to determine this parameter both experimentally[10-12] and theoretically.[13-15] However, these works gave quite different values. For example, Hong *et al.* measured the valence band offset ($\Delta E_{VBO}$) at a ZnO/GaN (0001) heterointerface by means of ex situ ultraviolet and x-ray photoemission spectroscopy (UPS/XPS). The $\Delta E_{VBO}$ was estimated to be 0.8 eV - 1.0 eV in their work[10]. Veal *et al.* evaluated the $\Delta E_{VBO}$ of ZnO/GaN heterojunctions using the transitivity rule of the natural band offsets between III-V compounds and II-IV compounds.[16] Based on their XPS data, they gave a $\Delta E_{VBO}$ value of 1.37 eV.[12] Nakayama *et al.* calculated the $\Delta E_{VBO}$ at ZnO/GaN interface using density-functional theory (DFT) calculations within local density approximation (LDA).[13] The $\Delta E_{VBO}$ was estimated to be around 1.6 eV. However, the *p-d* repulsive interactions were not taken into account in their work. Pezold *et al.* adopted a generalized gradient approximation (GGA) in the form of Perdew-Burke-Ernzerhof (PBE) for the exchange-correlation functional in evaluating the $\Delta E_{VBO}$ values at the cation- and anion-compensated ZnO/GaN interfaces, and predicted that they are 1.0 eV and 0.5 eV, respectively.[14] By taking the Coulomb repulsive interaction of 3*d* electrons into account in the GGA+U strategy, Huda *et al.* obtained a $\Delta E_{VBO}$ value of 0.7 eV for ZnO/GaN nonpolar ($1\bar{1}00$) interface.[15] In our previous work, we adopted the Perdew-Wang (PW91) XC functional in the GGA+U strategy and got a $\Delta E_{VBO}$ of 1.04 eV for the same interface.[9] Obviously, the $\Delta E_{VBO}$ values given by DFT calculations are sensitive to the employed exchange-correlation functionals. However, both LDA and GGA functionals underestimated the band gaps of *w*-ZnO and *w*-GaN. Although, this can be partially overcome by using LDA+U or GGA+U strategy, the calculated band gaps are still much lower than the experimental values.[9, 15] Therefore, the $\Delta E_{VBO}$ values given by using these functionals remain doubtful.

Additionally, the band gaps of *w*-ZnO and *w*-GaN crystals correspond to adsorption peaks at the ultraviolet region. Solar harvesting applications require strong adsorption in visible light region. Tuning the band gap of these heterostructures becomes quite crucial. The (Ga$_{1-x}$Zn$_x$)(N$_{1-x}$O$_x$) solid solution formed in the region near the ZnO/GaN interfaces is expected to be a promising candidate to reach this goal. Therefore, band gap calculations of this solid solution and the $\Delta E_{VBO}$ values at the ZnO/(Ga$_{1-x}$Zn$_x$)(N$_{1-x}$O$_x$) and (Ga$_{1-x}$Zn$_x$)(N$_{1-x}$O$_x$)/GaN interfaces at high level are desirable.

Here, we report our DFT calculations on the band alignment of ZnO/GaN and ZnO/(Ga$_{1-x}$Zn$_x$)(N$_{1-x}$O$_x$)/GaN heterostructures using a Heyd-Scuseria-Ernzerhof (HSE) hybrid functional. We found that the optimized HSE functional reproduces well the band gaps of *w*-ZnO and *w*-GaN crystals. In contrast, GGA and GGA+U schemes underestimated not only the band gaps of the two crystals but also the $\Delta E_{VBO}$ value at the interface between them. The $\Delta E_{VBO}$ at a ZnO/GaN $(1\bar{1}00)$ hetero-interface given by the HSE functional (1.588 eV) is much higher than that given by GGA and GGA+U schemes (0.7 eV -1.04 eV). Our calculations also indicate that the $(Ga_{1-x}Zn_x)(N_{1-x}O_x)$ ($x$=0.125) solid solution has a direct band gap of about 2.608 eV and adsorption peaks at the visible light region, in good agreement with the experimental data. The effective masses of the carriers in the solid solution are comparable to those of the host materials. More interestingly, the $(Ga_{1-x}Zn_x)(N_{1-x}O_x)$ solid solution can form type-II band alignment with both ZnO and GaN crystals. This suggests that the formation of $(Ga_{1-x}Zn_x)(N_{1-x}O_x)$ solution at the ZnO/GaN interfacial region of core/shell *h*-NWs can improve the visible light adsorption ability and carrier collection efficiency, which are crucial for the design of next-generation solar cells.

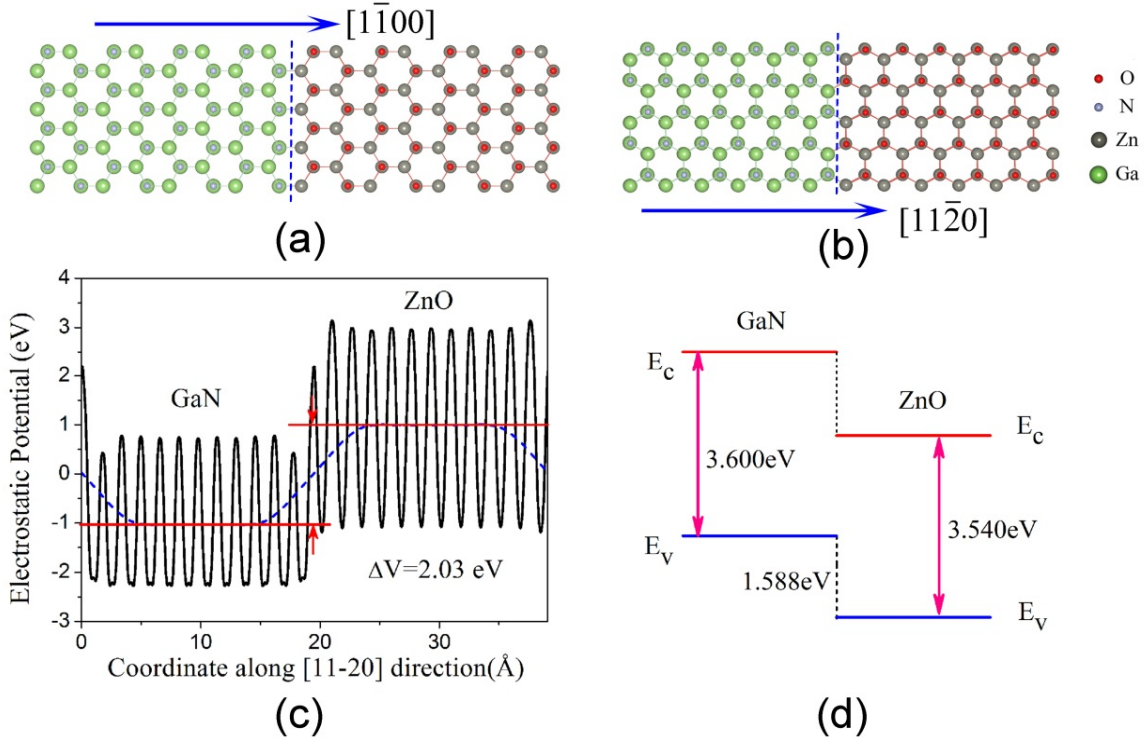

FIG. 1. (Color online) The slab models of ZnO/GaN (a) $[1\bar{1}00]$ and (b) $[11\bar{2}0]$ interfaces. (c) Electrostatic potential profile for $[11\bar{2}0]$ slab interface. (d) ZnO/GaN type-II band-offset schematic illustration obtained by *mod*-HSE06 calculations.

Our DFT calculations described in this work were performed using the projected augmented wave (PAW) basis as implemented in the Vienna ab initio simulation package (VASP).[17-19] The *3d* electrons of Zn and Ga ions are treated as valence electrons. A plane-wave cutoff energy of 450 eV was employed, and the atomic positions were optimized using the conjugate gradient (CG) scheme without any symmetric restrictions until the maximum force on each of them was less than 0.01 eV/Å. GGA(PBE)+U scheme was employed for the structural optimization. To improve the accuracy of electronic structure calculations A more accurate hybrid functional (HSE) presented by Heyd, Scuseria, and Ernzerhof[20, 21] was adopted in this work. A screened short-range Hatree-Fock (HF) exchange term is included in the HSE functional. The fraction of the HF exchange term is represented by an empirical parameter, $\alpha$, which is 0.25 for most semiconducting materials. In this work, we used optimized $\alpha$ values of 0.375 for *w*-ZnO and 0.25 for *w*-GaN (denoted as *mod*-HSE06 hereafter). The calculated band gaps are 3.56 eV (*w*-ZnO) and 3.60 eV (*w*-GaN), in good agreement with experimental data, 3.37 eV and 3.44 eV. We also employed a GGA+U strategy to make a comparison study. The U values representing the Coulomb repulsive interaction were set to 4.7 eV and 5.4 eV for the *3d* electrons of Zn and Ga atoms, which coincide with experimental data.[22, 23] The band gaps obtained from the GGA+U strategy are 1.48 eV for *w*-ZnO and 2.95 eV for *w*-GaN, much smaller than the experimental values.

On the basis of the electron density given by the DFT calculations, the electrostatic potential $V(\vec{r})$ can be calculated by solving Poisson equation. The planar-averaged potential $\bar{V}(z)$ across the ZnO/GaN interface was then obtained using the expression:

$$\bar{V}(z) = \frac{1}{S}\int_S V(\vec{r})dxdy \qquad (1)$$

where $S$ represents the area of a unit cell in the plane parallel to the interface (*xy*-plane). The macroscopic average $\bar{\bar{V}}(z)$ is accomplished by averaging $\bar{V}(z)$ at each point over a distance corresponding to one period ($L$),

$$\bar{\bar{V}}(z) = \frac{1}{L}\int_{-L/2}^{L/2}\bar{V}(z')dz' \qquad (2)$$

The valence band offset $\Delta E_{VBO}$ at ZnO/GaN heterointerface can be evaluated using the equation:

$$\Delta E_{VBO} = \Delta\bar{\bar{V}}|_{ZnO/GaN} + (E_{VBM} - \bar{\bar{V}})|_{ZnO} - (E_{VBM} - \bar{\bar{V}})|_{GaN} \qquad (3)$$

TABLE I. The band gaps, valence band offsets ($\Delta E_{VBO}$), and each component of the valence band offsets in Eq. (3) obtained by using different functionals. The energies are in eV.

| | Band Gap | | $\Delta \overline{\overline{V}}(z)|_{ZnO/GaN}$ | $(E_{VBM} - \overline{\overline{V}})|_{ZnO}$ | $(E_{VBM} - \overline{\overline{V}})|_{GaN}$ | $\Delta E_{VBO}$ |
| --- | --- | --- | --- | --- | --- | --- |
| | ZnO | GaN | | | | |
| mod-HSE06 | 3.540 | 3.600 | 2.000 | -0.330 | 3.258 | 1.588 |
| PBE+U | 1.481 | 2.580 | 2.000 | 1.063 | 3.808 | 0.745 |
| PW91+U | 1.415 | 2.163 | 1.858 | 0.746 | 3.644 | 1.040 |
| PBE | 0.779 | 1.859 | 2.000 | 0.847 | 3.302 | 0.455 |
| PW91 | 0.835 | 1.861 | 1.858 | 0.778 | 3.267 | 0.631 |

where the first term represents the $\overline{\overline{V}}(z)$ difference between the two components in ZnO/GaN heterostructure, and the last two terms are the difference between the valence band minimum (VBM) energy ($E_{VBM}$) and $\overline{\overline{V}}(z)$ of the corresponding isolated component in bulk crystal. Such strategy has been widely adopted in calculations of $\Delta E_{VBO}$ at nonpolar interfaces.[13,14,24,25]

We first calculated the $\Delta E_{VBO}$ values at a nonpolar $(1\bar{1}00)$ ZnO/GaN heterointerface using a slab model, as shown in FIG. 1. The supercell has eight ZnO bilayers and eight GaN bilayers with minimum number of "wrong bonds" at the interface. The $\Delta E_{VBO}$ values at the nonpolar interface obtained by means of different exchange-correlation (XC) functionals are listed in TABLE I. Obviously, $\Delta E_{VBO}$ is sensitive to the adopted XC functionals. The mod-HSE06 functional gives the highest $\Delta E_{VBO}$ value (1.588 eV). This suggests that GGA functional underestimates not only the band gaps but also the $\Delta E_{VBO}$ values. Although such underestimation can be partially overcome by taking the Coulomb repulsive interaction (U) of 3d electrons into account in the GGA+U scheme, the $\Delta E_{VBO}$ is still much lower than that obtained from the mod-HSE06 functional.

Further analysis indicates that the sensitivities of the three terms in Eq. (3) to the functionals are different. The first term $\Delta \overline{\overline{V}}|_{ZnO/GaN}$ which is determined by the electron density distribution across the interface is nearly independent of the adopted functional, except the PW91 functional. This implies that these functionals can give reasonable electron density and thus reliable electrostatic potential. However, the energies of the VBM of isolated w-ZnO and w-GaN crystals given by GGA or GGA+U are questionable, since neither of them can reproduce the band gaps of these crystals. Scissors operator approximation based on GGA or GGA+U calculations doesn't work in determining the band alignment of heterostructured materials. Therefore, the DFT calculations under the precision of HSE or GW are needed not only for evaluating the band gaps but also for determining the band alignment.

We then calculated the electronic structures of ZnO/GaN solid solution using the mod-HSE06 functional. The optimized atomic structure of $(Ga_{1-x}Zn_x)(N_{1-x}O_x)$ with $x=0.125$, corresponding to one Ga-N pair of a 2×2×1 w-GaN supercell being substituted by one Zn-O pair is shown in FIG. 2(a) and (b). Thanks to the slight lattice mismatch between w-GaN and w-ZnO (~1.86%), the relaxed lattice parameter of the solid solution is essentially unchanged and the calculated formation energy with respect to w-GaN and w-ZnO is almost neglectable. The α value in mod-HSE06 functional is determined using this principle: $\alpha = x\alpha_1 + (1-x)\alpha_2$, where $\alpha_1$ and $\alpha_2$ are set to the values optimized for w-ZnO and w-GaN crystals[26]. The band structure of $(Ga_{0.875}Zn_{0.125})(N_{0.875}O_{0.125})$ is figured out as FIG. 2(c). There is a direct band gap of 2.608 eV at the Γ point, in good agreement with the experimental value.[27]

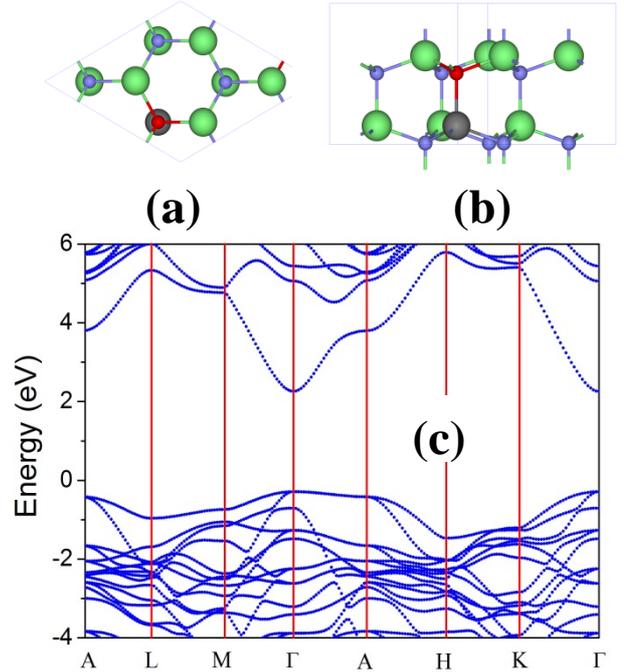

FIG. 2. (Color online) (a) top and (b) side views of $(Ga_{1-x}Zn_x)(N_{1-x}O_x)$ ($x=0.125$) unit cell. The green, gray, blue and red balls represent the Ga, Zn, N and O atoms, respectively. (c) Electronic band structures of $(Ga_{0.875}Zn_{0.125})(N_{0.875}O_{0.125})$ solid solution, calculated by using mod-HSE06.

To reveal the origins of the narrow band in $(Ga_{0.875}Zn_{0.125})(N_{0.875}O_{0.125})$, we plot the wavefunctions of the VBM and the conduction band minimum (CBM) at Γ point, as shown in FIG. 3(a),(b) and (c). It is clear that the VBM states are contributed mainly by the N-*2p* and Zn-*3d* atomic orbitals, while CBM comes mainly from the Ga-*4s* and Ga-*4p* atomic orbitals. This is also consistent with the partial density of states shown in FIG. 3(d). The strong *p-d* exchange interaction between N-*2p* and Zn-*3d* shifts the VBM upward, while the CBM arising from Ga-*4s* and Ga-*4p* states remains unchanged, making the band gap of $(Ga_{1-x}Zn_x)(N_{1-x}O_x)$ solid solution narrower than those of the host materials. With the increase of ZnO concentration $x$, the interaction between N-*2p* and Zn-*3d* increases, and the band gap of the solid solution decreases. This trend is in good agreement with the experimental findings.[27] The tunable band gaps of $(Ga_{1-x}Zn_x)(N_{1-x}O_x)$ are suitable for overall water splitting under visible-light irradiation.[4, 27]

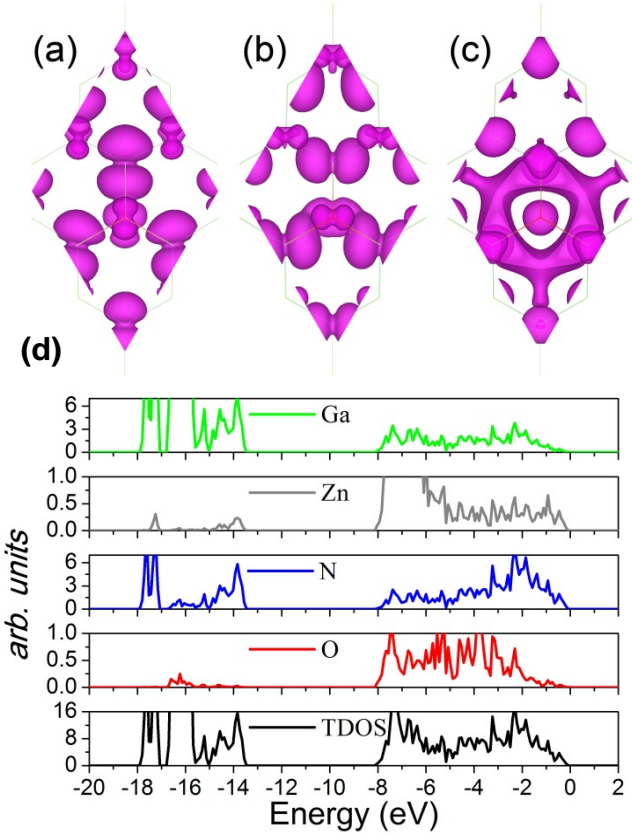

FIG. 3. (Color online) Isosurfaces of the Kohn-Sham states: (a and b) the valence band maximum (VBM) ; (c) the conduction band minimum (CBM) at Γ point of $(Ga_{1-x}Zn_x)(N_{1-x}O_x)$ ($x$=0.125). The isovalue is 0.0025Å$^{-3/2}$. (d) The PDOS and TDOS of $(Ga_{1-x}Zn_x)(N_{1-x}O_x)$ ($x$=0.125), obtained by projecting the total density of states onto the four type atoms, respectively. The energy at Fermi level was set to zero.

Carrier mobility in semiconductors is closely related to their effective masses (*EMs*), i.e. the small *EM* corresponds to high mobility. We calculated the *EMs* of *w*-ZnO, *w*-GaN, and $(Ga_{0.875}Zn_{0.125})(N_{0.875}O_{0.125})$ solid solution using different XC funtionals. For simplification, we regard the $k$ dependence of the Kohn-Sham eignenergy around the conduction band bottom as a parabolic dependence and ignore the cross term of the *EMs* such as $m_{k_z}$:

$$e(\vec{k}) = \frac{\hbar^2}{2}\left(\frac{k_x^2}{m_{k_x}} + \frac{k_y^2}{m_{k_y}} + \frac{k_z^2}{m_{k_z}}\right) \quad (4)$$

where $k_x$ is defined as the direction from Γ to M points in the $k$ space of the first Brillouin zone. $k_z$ is the direction from Γ to A points, while $k_y$ is perpendicular to $k_x$ and $k_z$. The calculated *EM* values by means of GGA+U and *mod*-HSE06 functionals are listed in TABLE II. Obviously, GGA+U calculations underestimate the *EMs* of *w*-ZnO and *w*-GaN crystals, compared to the experimental data which are 0.29 $m_0$[28] and 0.20 $m_0$[29] for *w*-ZnO and *w*-GaN, respectively ($m_0$ is the electron rest mass), corresponding to $m_{k_x}$ or $m_{k_y}$. The *EM* values given by *mod*-HSE06 functionals are comparable to the results of accurate quasiparticle calculations under GW approximation (GWA)[30] and experimental data. This implies that the *mod*-HSE06 functionals can reproduce not only the band gaps but also the band dispersion of *w*-ZnO and *w*-GaN crystals. The *EM* values of $(Ga_{0.875}Zn_{0.125})(N_{0.875}O_{0.125})$ is very close to those of *w*-GaN and smaller than those of *w*-ZnO, suggesting that the carriers in this solid solution have high mobility.

TABLE II. The effective masses of *w*-ZnO, *w*-GaN and their solid solution $(Ga_{1-x}Zn_x)(N_{1-x}O_x)$ with $x$= 0.125 calculated from GGA+U and *mod*-HSE06, respectively. The unit is the electron rest mass $m_0$.

| Material | *w*-ZnO | | *w*-GaN | | $(Ga_{1-x}Zn_x)(N_{1-x}O_x)$ | |
|---|---|---|---|---|---|---|
| | GGA+U | HSE | GGA+U | HSE | GGA+U | HSE |
| $m_{k_x}(m_{k_y})$ | 0.161 | 0.276 | 0.186 | 0.219 | 0.175 | 0.202 |
| $m_{k_z}$ | 0.145 | 0.255 | 0.159 | 0.187 | 0.166 | 0.194 |

The optical adsorption properties of *w*-GaN, *w*-ZnO and their solid solution $(Ga_{0.875}Zn_{0.125})(N_{0.875}O_{0.125})$ were then investigated by computing the complex dielectric function, $\varepsilon(\omega)=\varepsilon_1(\omega)+i\varepsilon_2(\omega)$, in which the imaginary part $\varepsilon_2(\omega)$ reflects optical abosorption at a given frequency $\omega$. The imaginary part $\varepsilon_2(\omega)$ can be written[31] as

$$\varepsilon_{ab}^{(2)}(\omega) = \frac{4\pi^2 e^2}{\Omega}\lim_{q\to 0}\frac{1}{q^2}\sum_{c,v,k}2w_k\delta(e_{ck}-e_{vk}-\omega)\times\langle u_{ck+e_a q}|u_{vk}\rangle\langle u_{ck+e_b q}|u_{vk}\rangle^* \quad (5)$$

where the indices $c$ and $v$ refer to conduction and valence band states respectively which were determined by using the *mod*-HSE06 functional, and $u_{ck}^{\vec{r}}$ is the cell periodic part of the wavefunctions at the *k*-point. The knowledge of $\varepsilon_2(\omega)$ over a wide frequency range allows one to obtain $\varepsilon_1(\omega)$ using the Kramers-Kronig relation. The calculated $\varepsilon_2(\omega)$ along [0001] direction for these three materials is shown in FIG. 4. It is clear that there are two obvious absorption peaks at about 2.6 eV and 2.9 eV for the solid solution compared with those of the host materials at about 3.4 eV. The appearance of adsorption peaks at the visible light region is quite curial for the applications of this material in solar harvesting.

Finally, we calculated the band alignment in ZnO/$(Ga_{1-x}Zn_x)(N_{1-x}O_x)$/GaN heterostructure using above mentioned strategy. The slab models employed in these calculations are shown in FIG. 5(a) and (b). Our calculations indicate that the $DE_{VBO}$ values at ZnO/(Ga$_{0.875}$Zn$_{0.125}$)(N$_{0.875}$O$_{0.125}$) and (Ga$_{0.875}$Zn$_{0.125}$)(N$_{0.875}$O$_{0.125}$)/GaN nonpolar $(1\bar{1}00)$ interfaces are 0.997 eV and 0.592 eV, respectively. It is interesting to see that both interfaces in the ZnO/(Ga$_{1-x}$Zn$_x$)(N$_{1-x}$O$_x$)/GaN heterostructure exhibit type-II band alignment features, as shown in FIG. 5(c). Type-II band alignment is also accompanied by the natural spatial charge separation behaviors[5, 32, 33] at the interfaces. Therefore, the formation of (Ga$_{1-x}$Zn$_x$)(N$_{1-x}$O$_x$) in the region near the ZnO/GaN interface is advantageous not only for enhancing the visible light adsorption ability and for improving the carrier collection efficiency, both of which are the goals of the next-generation solar cells.

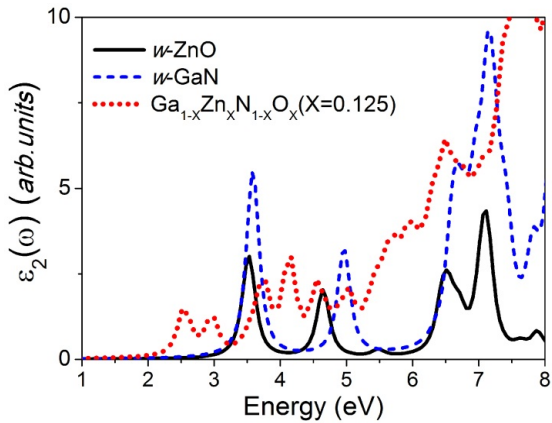

FIG. 4. (Color online) Imaginary parts of dielectric constant for wurtzite ZnO, GaN and (Ga$_{1-x}$Zn$_x$)(N$_{1-x}$O$_x$) ($x$=0.125) respectively.

On the basis of these results, we propose a core/intermediate/shell solar cell architecture model composing of ZnO/(Ga$_{1-x}$Zn$_x$)(N$_{1-x}$O$_x$)/GaN heterostructured nanowires, as shown in the inset of FIG. 5. Different from the ZnO/GaN core-shell model, we introduce (Ga$_{1-x}$Zn$_x$)(N$_{1-x}$O$_x$) solid solution to the nanowires as intermediate region to increase the visible light adsorption ability due to the tunable band gap. The type-II band alignment at the two interfaces in this model and the high carrier mobility facilitate the spatial separation of the carriers generated in the intermediate region. The overall efficiency of the solar cell model is therefore expected to be significantly higher than the ZnO/GaN core-shell model. Of course, the formation of (Ga$_{1-x}$Zn$_x$)(N$_{1-x}$O$_x$) solid solution in the heterostructured nanowires becomes an open question for the experimentalists.

In summary, our DFT calculations show the conventional GGA (or GGA+U) functionals underestimate the band gaps, electron effective masses of $w$-ZnO and $w$-GaN, and the valence band offset between these two crystals, whereas those obtained from the *mod*-HSE06 functional are comparable to the experimental data. The (Ga$_{1-x}$Zn$_x$)(N$_{1-x}$O$_x$) solid solution with $x$=0.125 has a direct band gap of 2.608 eV at the Γ point and strong light adsorption peaks in visible light region, arising from the strong interaction between N-$2p$ and Zn-$3d$ states. Both ZnO/(Ga$_{0.875}$Zn$_{0.125}$)(N$_{0.875}$O$_{0.125}$) and (Ga$_{0.875}$Zn$_{0.125}$)(N$_{0.875}$O$_{0.125}$)/GaN nonpolar interfaces have type-II band alignment features, accompanied by spatial charge separation behaviors. Based on these results, we propose that the formation of (Ga$_{1-x}$Zn$_x$)(N$_{1-x}$O$_x$) solid solution in the region between GaN and ZnO components of GaN/ZnO core-shell heterostructured nanowires can improve the visible light adsorption ability and carrier collection efficiency, which are crucial for the design of next-generation solar cells.

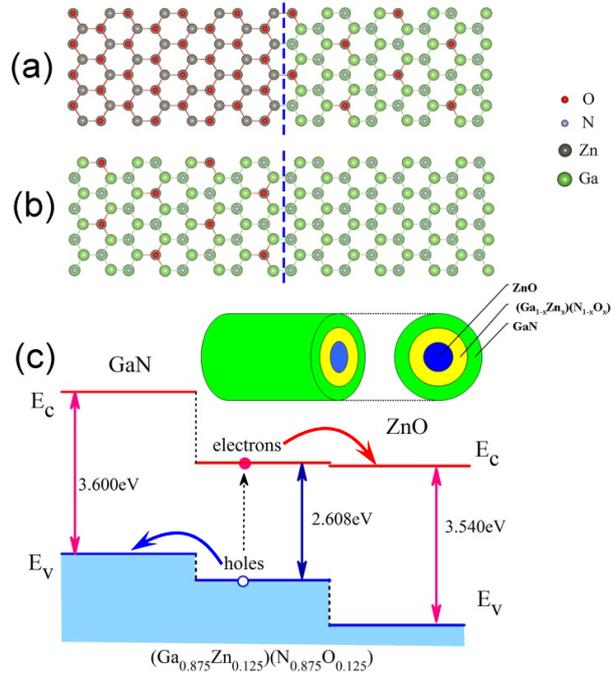

FIG. 5. (Color online) The slab models of nonpolar $(1\bar{1}00)$ (a) ZnO/(Ga$_{0.875}$Zn$_{0.125}$)(N$_{0.875}$O$_{0.125}$) and (b) (Ga$_{0.875}$Zn$_{0.125}$)(N$_{0.875}$O$_{0.125}$)/GaN interfaces; (c) the sketch representation of sandwich structure of ZnO/(Ga$_{0.875}$Zn$_{0.125}$)(N$_{0.875}$O$_{0.125}$)/GaN core/intermediate/shell NWs and its band alignment.

This work is supported by the National Basic Research Program of China (No. 2012CB932302), the National Natural Science Foundation of China (No. 10974119), the Natural Science Fund for Distinguished Young Scholars of Shandong Province (No. JQ201001), and the Graduate Independent Innovation Foundation of Shandong University (GIIFSDU, No. YZC09075).